\documentclass[twocolumn]{jpsj2} %% two-column layout

\title{Numerical calculation of the Landauer conductance
through an interacting electron system
in the Hartree-Fock approximation}

\author{Yoichi \textsc{Asada}}
\inst{
Department of Physics, Tokyo Institute of Technology, 
2-12-1 Ookayama, Meguro-ku, Tokyo 152-8551, Japan}

\date{\today}

\abst{
We develop a new numerical method to calculate the Landauer conductance
through an interacting electron system in the first order perturbation
or in the self-consistent Hartree-Fock approximation. It is applied to
one and two dimensional systems with nearest-neighbor electron-electron
interaction.
}
\kword{Landauer conductance, electron-electron interaction,
Hartree-Fock approximation}

\begin{document}
\maketitle

\section{Introduction}
\label{sec:intro}
The interplay between disorder and electron-electron interaction
in quantum electron transport phenomena is one of the most challenging
problems. Although the non-interacting approximation successfully
explains many aspects of experiments \cite{kramer:93review}, we may
need to take into account the electron-electron interaction to
understand some phenomena, such as the metallic behavior in two
dimensional (2D) systems in Si-MOS and heterostructures,
\cite{abrahams:01} and the critical phenomena of the 3D metal-insulator
transitions.\cite{itoh:04} In theoretical works, important corrections due
to electron-electron interaction in disordered electron systems were
found.\cite{fukuyama:80,fukuyama:81,altshuler:85review,efros:75,efros:76}
Furthermore, the study of the nonlinear $\sigma$ model suggested new
universality classes for the metal-insulator transition. \cite{belitz:94}

The Hartree-Fock (HF) approximation was employed in some
numerical works to study interacting disordered electron systems.
\cite{macdonald:86,yang:93,yang:95,epperlein:97,jeon:99-1,jeon:99-2,
levit:99,lee:02,heidarian:04}
It enables us to simulate relatively large systems.
Even at the level of the HF approximation, however,
our understanding is not yet complete.
One of the reasons is the lack of numerical simulations
of the Landauer conductance. As used in the
scaling theory of Anderson localization \cite{abrahams:79}
(See Refs.~\citen{shapiro:87,slevin:01-1,slevin:03-1}
for more detailed discussions on the scaling hypothesis
of the Landauer conductance.),
the conductance is one of the most important physical quantities
to characterize a disordered electron system. We expect that
numerical simulation of the Landauer conductance would improve our
understanding on the interplay between disorder and electron-electron
interaction.

Motivated by this, we have decided to perform a numerical calculation
of the Landauer conductance in interacting disordered electron systems
in the HF approximation. As a first step toward it we have developed a
numerical method, which we report here.

The generalization of the Landauer approach to interacting electron
systems is a very active topic not only to study interacting
disordered electron systems but also to study transport phenomena
through low dimensional electron systems.
Although much progress has been made in the generalization
of the Landauer approach,
\cite{datta,haug,ng:88,meir:92,izumida:97,kawabata:98,oguri:97,oguri:99,
tanaka:02,oguri:05,sushkov:01,molina:03,molina:04-1,meden:03}
the calculation method is still being improved.

This paper is organized as follows:
In \S~\ref{sec:model} the model considered is described and
in \S~\ref{sec:landauer} the Landauer formula is described.
The HF approximation is described in \S~\ref{sec:hartreefock}.
In \S~\ref{sec:wideband}, we explain a new numerical method, which
we call wide band method. In \S~\ref{sec:result1d} and
\S~\ref{sec:result2d}, we apply the wide band method to 1D and 2D systems
of interacting electrons. The last section is devoted to summary and
discussion.

\section{Model}
\label{sec:model}
\begin{figure}[tb]
\includegraphics[width=\linewidth]{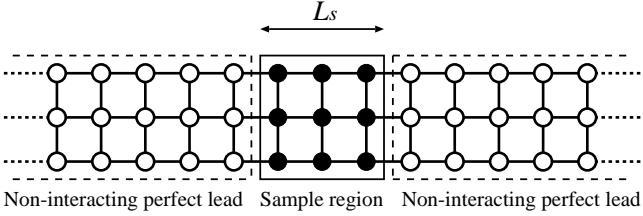}
\caption{An example of the Landauer geometry
to calculate the two terminal conductance.
Two non-interacting perfect leads are attached to the sample.}
\label{fig:landauersystem}
\end{figure}

We consider spinless electrons on a 2D square lattice. As illustrated
in Fig.~\ref{fig:landauersystem}, the system considered consists of
a sample region with size $L_s\times L_y$ (denoted by $\cal S$)
and two semi-infinite leads with width $L_y$ (denoted by $\cal L$).
We impose fixed boundary conditions in the transverse direction.
We suppose that electrons are interacting in the sample region, while
non-interacting in the lead region. In the present paper we do not
consider random potential for simplicity. We take the $x$ direction
as the current direction and $y$ direction as the transverse direction.

The tight binding Hamiltonian is given by
\begin{eqnarray}
 {\cal H}&=&{\cal H}_s+{\cal H}_{\ell}+{\cal H}_{\ell s}+{\cal H}_u,
 \label{eq:horg}
\end{eqnarray}
where ${\cal H}_s$ is the non-interacting part for the sample region
$\cal S$, ${\cal H}_{\ell}$ the Hamiltonian for the perfect leads
at the left and right, ${\cal H}_{\ell s}$ the coupling between
the sample and leads, and ${\cal H}_u$ electron-electron interaction
in the sample region. They are given by
\begin{eqnarray}
 {\cal H}_s&=&
-t_s\sum_{\langle i,j \rangle (i,j\in \cal S)} c_i^{\dagger}c_j,
 \label{eq:hs}
 \\
 {\cal H}_{\ell}&=&-t_{\ell} \sum_{\langle i,j\rangle (i,j\in {\cal L})}
 c_i^{\dagger}c_j,
 \label{eq:hl}
 \\
 {\cal H}_{\ell s}&=&-t_{\ell s}\sum_{y=1}^{L_y}
 \left( c_{0,y}^{\dagger}c_{1,y}+
 c_{1,y}^{\dagger}c_{0,y} \right)\nonumber \\
 &&-t_{\ell s}\sum_{y=1}^{L_y}\left(c_{L_s, y}^{\dagger}c_{L_s+1, y}
 +c_{L_s+1, y}^{\dagger}c_{L_s, y}\right),
 \label{eq:hls}
 \\
 {\cal H}_u&=&\frac{1}{2}\!\sum_{i,j (i,j \in {\cal S}, i\neq j)}\!\!
\left(c_i^{\dagger}c_i\!-\!K \right)\!U_{i,j}\!
\left(c_j^{\dagger}c_j\!-\!K\right). \label{eq:h_int}
\end{eqnarray}
Here $c_i^{\dagger}(c_i)$ denotes the creation (annihilation) operator
of an electron at the site $i$, $U_{i,j}$ is the interaction between
electrons at the sites $i$ and $j$, and $K$ is a positive uniform charge.
Hopping is restricted to nearest-neighbors.
We suppose that the hopping parameters $t_s$, $t_{\ell}$, and
$t_{\ell s}$ are real, so the system has time reversal symmetry.

\section{Landauer conductance}
\label{sec:landauer}
The Landauer conductance $g$ at zero temperature in the linear
response regime is expressed using Green's functions as
\begin{equation}
 g=\frac{e^2}{h}{\rm tr} \left[ \Gamma^{\rm (L)}(\mu)
 G_{1L_s}^a(\mu) \Gamma^{\rm (R)}(\mu) G_{L_s 1}^r(\mu)
 \right].
 \label{eq:landauerconductance}
\end{equation}
Here $\mu$ is the chemical potential of the system, $G_{L_s 1}^r$
and $G_{1L_s}^a$ the submatrices of the retarded and advanced
Green's functions in equilibrium, and $\Gamma^{\rm (L,R)}$ the matrices
determined by the attached leads. \cite{datta}
The matrices $\Gamma^{\rm (L,R)}$ are explicitly shown in
Appendix \ref{app:green}.

It has been shown in Refs.~\citen{oguri:97} and
\citen{tanaka:02} that the expression
(\ref{eq:landauerconductance}), which is well known for non-interacting
electrons, \cite{datta} is valid even when electrons are interacting
in the sample region at zero temperature. Within the HF approximation
the expression (\ref{eq:landauerconductance}) can also be justified
based on the Keldysh Green's function method. \cite{datta}

In the following sections, we only consider the retarded Green's
function $G^r$ since the advanced Green's function $G^a$
is simply Hermitian conjugate of the retarded Green's function.

\section{Hartree-Fock approximation}
\label{sec:hartreefock}

We describe the HF approximation in the Green's function formalism
\cite{fetter} for the system (\ref{eq:horg}).

The Dyson equation for $G^r$ in the HF approximation is obtained by
expanding the time ordered Green's function in ${\cal H}_u$ and by
performing an analytic continuation. We have
\begin{equation}
 G^r(\epsilon)=G_0^r(\epsilon)+G_0^r(\epsilon)
 \Sigma^{(\rm HF)} G^r(\epsilon).
 \label{eq:dysonSCHF}
\end{equation}
Here $G_0^r$ is the retarded Green's function for the non-interacting
part of the Hamiltonian
${\cal H}_0={\cal H}_s+{\cal H}_{\ell}+{\cal H}_{\ell s}$,
and $\Sigma^{\rm (HF)}$ is the self-energy due to electron-electron
interaction in the HF approximation, which is non-zero only in the
sample region. It consists of Hartree and exchange contributions:
\begin{eqnarray}
 \Sigma^{\rm (HF)}_{i,i}=\sum_{j (j\neq i)}U_{i,j}
 \left[-
 \frac{1}{\pi}\int_{-\infty}^{\mu} \mathrm{d}\epsilon
 {\rm Im}G^r_{j,j}(\epsilon)-K \right],
 \label{eq:hartree}
\end{eqnarray}
\begin{eqnarray}
 \Sigma^{\rm (HF)}_{i,j(i\neq j)}=
 \frac{U_{i,j}}{\pi} \int_{-\infty}^{\mu} \mathrm{d}\epsilon
 {\rm Im}G^r_{i,j}(\epsilon),
 \label{eq:exchange}
\end{eqnarray}
where $i,j\in {\cal S}$. (In more general expression, ${\rm Im} G^r(\epsilon)$
is replaced with
$(-\mathrm{i}/2)\left[G^r(\epsilon)-G^a(\epsilon)\right]$.
For our model, they are the same.) From (\ref{eq:dysonSCHF}), we have
\begin{equation}
 G^r(\epsilon)=
 \left[\epsilon-H_{\rm 0}
-\Sigma^{\rm (HF)}+\mathrm{i}\eta\right]^{-1}.
 \label{eq:Gschf}
\end{equation}
Here $H_{\rm 0}$ is the single particle Hamiltonian in the matrix form
corresponding to the non-interacting part ${\cal H}_0$ and $\eta$ is an
infinitesimal positive number. The HF self-energy $\Sigma^{\rm (HF)}$
is a solution of the self-consistent equations
(\ref{eq:hartree})--(\ref{eq:Gschf}).

In practice there are two difficulties we need to solve:
\begin{itemize}
\item{
Since the Landauer geometry corresponds to an open system, the size
of the matrix $H_0$ is infinite. We need to make the matrix size
finite to perform numerical simulations.
}
\item{
To calculate the HF self-energy $\Sigma^{\rm (HF)}$, we need
to perform an integral over $\epsilon$.
}
\end{itemize}

The first problem can be solved by taking account of the effects of
the semi-infinite leads in terms of a self-energy.
\cite{datta,ando:91,moldoveanu:01,agarwal:06}
For example, if we expand the Green's function in ${\cal H}_{\ell s}$ in
addition to ${\cal H}_u$, we obtain the Dyson equation for the retarded
Green's function $G_{i,j}^r$ $(i,j\in {\cal S})$ in the HF approximation
in a finite size matrix form that is closed in the sample region $\cal S$.
The Green's function $G_{i,j}^r$, with $i,j \in {\cal S}$, is written as
\begin{eqnarray}
  G^r(\epsilon)=
  \left[\epsilon-H_{s}-\Sigma^{(\ell)r}(\epsilon)
 -\Sigma^{({\rm HF})}+\mathrm{i}\eta
  \right]^{-1}.
 \label{eq:Gschffinite}
\end{eqnarray}
Here $H_{s}$ is the single particle Hamiltonian in the matrix form
corresponding to ${\cal H}_s$,
$\Sigma^{\rm (HF)}$ is the HF self-energy which is given by
Eqs.~(\ref{eq:hartree}) and (\ref{eq:exchange}), and
$\Sigma^{(\ell)r}$ is the retarded self-energy due to the attached
semi-infinite leads. An element of the self-energy
$\Sigma_{xy,x'y'}^{(\ell)r}$ is non-zero only when $x=x'=1$ or $x=x'=L_s$.
The non-zero elements are written as (see Appendix \ref{app:green})
\begin{eqnarray}
 \Sigma^{(\ell)r}_{Xy,Xy'}(\epsilon)
= \frac{t_{\ell s}^2}{t_{\ell}}
\sum_{n} \chi_n(y) \chi_n(y')
\zeta\left(
\frac{\epsilon}{t_{\ell}}-\lambda_n
\right),
\label{eq:sigmaleadorg}
\end{eqnarray}
where $X=1$ or $L_s$.
The Green's function $G_{i,j}^{r}$ ($i,j\in {\cal S}$) of
(\ref{eq:Gschffinite}) is exactly the same as that of (\ref{eq:Gschf})
since we have taken into account all orders in ${\cal H}_{\ell s}$.
Now in (\ref{eq:Gschffinite}) the size of the matrices is finite,
$L_sL_y\times L_sL_y$, so it is possible to perform numerical
calculations in principle.

As for the second problem, a numerical integration was employed
in similar approaches. \cite{moldoveanu:01,agarwal:06}
The numerical integration of the Green's function is not very
difficult if a simple system is considered. However the numerical
integration is troublesome in general, so we have developed
a method to avoid it. The method is explained in the next section.

\section{Wide band method}
\label{sec:wideband}

\begin{figure}[tb]
\includegraphics[width=\linewidth]{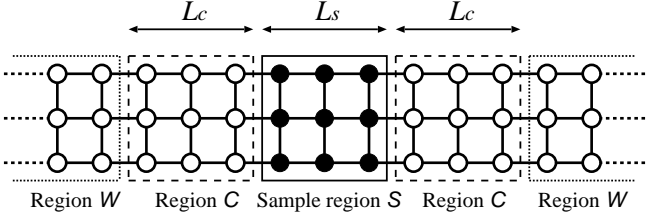}
\caption{The system envisaged in the wide band method.
Each non-interacting lead is
divided into two regions ${\cal C}$ and ${\cal W}$.}
\label{fig:widebandsystem}
\end{figure}

We have developed a new method, which we call wide band method,
to avoid the numerical integration as follows.

We change the system considered from the original Landauer system,
illustrated in Fig.~\ref{fig:landauersystem}, to another system
as illustrated in Fig.~\ref{fig:widebandsystem}. We call it
wide band system. The lead region
is divided into two parts: the region ${\cal C}$ up to a length $L_c$
on both sides of the sample, and the region ${\cal W}$ consisting
of two semi-infinite regions.  The wide band system is described by the
following Hamiltonian,
\begin{equation}
 \tilde{{\cal H}}={\cal H}_s+
 \tilde{{\cal H}}_{\ell}+{\cal H}_{\ell s}+{\cal H}_u.
 \label{eq:wbhamiltonian}
\end{equation}
Here ${\cal H}_s$, ${\cal H}_{\ell s}$, and ${\cal H}_u$ are already
given in (\ref{eq:hs}), (\ref{eq:hls}), and (\ref{eq:h_int}). The other
term $\tilde{{\cal H}}_{\ell}$ for the leads is made of three terms
\begin{eqnarray}
\tilde{{\cal H}}_{\ell}={\cal H}_c+{\cal H}_w+{\cal H}_{cw},
\end{eqnarray}
which are given by
\begin{eqnarray}
{\cal H}_c&=&-t_{\ell}\sum_{i,j (i,j \in {\cal C})}
c_i^{\dagger}c_j, \\
{\cal H}_w&=&\epsilon_w\sum_{i (i \in {\cal W})}c_i^{\dagger}c_i
-t_w\sum_{\langle i,j \rangle
(i,j\in {\cal W})} c_i^{\dagger} c_j,
\label{eq:hw}\\
{\cal H}_{cw}&=&-t_{cw}\sum_{y}
\left(
c_{X_{\mathrm{L}}-1,y}^{\dagger} c_{X_{\mathrm{L}},y}
+ c_{X_{\mathrm{L}},y}^{\dagger} c_{X_{\mathrm{L}}-1,y}
\right.\nonumber \\
&& \left.
+c_{X_{\mathrm{R}},y}^{\dagger} c_{X_{\mathrm{R}}+1,y}
+ c_{X_{\mathrm{R}}+1,y}^{\dagger} c_{X_{\mathrm{R}},y}
\right).
\label{eq:hlc}
\end{eqnarray}
with $X_{\mathrm{L}}=-L_c+1$ and $X_{\mathrm{R}}=L_s+L_c$. Here
${\cal H}_c$ is the Hamiltonian for the finite regions $\cal C$ on
the sample, ${\cal H}_w$ the Hamiltonian for the semi-infinite
regions $\cal W$, and ${\cal H}_{cw}$ the coupling between the regions
$\cal C$ and $\cal W$. Two hopping parameters $t_{w}, t_{cw}$ and
one parameter for uniform potential $\epsilon_w$ are introduced
for the wide band
system. If $t_w=t_{cw}=t_{\ell}$ and $\epsilon_w=0$,
the Hamiltonian (\ref{eq:wbhamiltonian}) is exactly the same as the
original Hamiltonian (\ref{eq:horg}), which is the system we want to
solve. However, as we explain below, we take a limit
$t_w, t_{cw}, \epsilon_w \rightarrow \infty$ under certain conditions
(conditions (\ref{eq:ratio1}) and (\ref{eq:ratio2})).
We call this limit wide band limit since
the band width in the region $\cal W$ becomes infinity
in the limit $t_w\rightarrow\infty$.

We expand the Green's functions in ${\cal H}_{cw}$ and ${\cal H}_u$.
The retarded Green's function for the wide band system
(\ref{eq:wbhamiltonian}) in the HF approximation is written as,
\begin{eqnarray}
  \tilde{G}^r(\epsilon)=
  \left[\epsilon-H_t-\tilde{\Sigma}^{(w)r}(\epsilon)
 -\tilde{\Sigma}^{\rm (HF)} +\mathrm{i}\eta
  \right]^{-1}.
 \label{eq:grWB1}
\end{eqnarray}
Here $H_t$ is the Hamiltonian in a matrix form corresponding to
${\cal H}_t={\cal H}_s+{\cal H}_{\ell s}+{\cal H}_c$,
$\tilde{\Sigma}^{\rm (HF)}$ the self-energy in the HF
approximation for the system (\ref{eq:wbhamiltonian})
\begin{equation}
 \tilde{\Sigma}^{\rm (HF)}_{i,i}=\sum_{j (j\neq i)}U_{i,j}
 \left[-
 \frac{1}{\pi}\int_{-\infty}^{\mu} \mathrm{d}\epsilon
{\rm Im}\tilde{G}^r_{j,j}(\epsilon)-K
 \right],
 \label{eq:hartreeWB}
\end{equation}
\begin{equation}
 \tilde{\Sigma}^{\rm (HF)}_{i,j(i\neq j)}=
 \frac{U_{i,j}}{\pi} \int_{-\infty}^{\mu} \mathrm{d}\epsilon
 {\rm Im}\tilde{G}^r_{i,j}(\epsilon),
 \label{eq:exchangeWB}
\end{equation}
where $i,j\in {\cal S}$,
and $\tilde{\Sigma}^{(w)r}$ the retarded
self-energy due to the semi-infinite region $\cal W$.
An element $\tilde{\Sigma}_{xy,x'y'}^{(w)r}$ is
non-zero only when $x=x'=X_{\mathrm{L}}$ or $x=x'=X_{\mathrm{R}}$.
The non-zero elements are given by
\begin{eqnarray}
 \tilde{\Sigma}^{(w)r}_{Xy,Xy'}(\epsilon)
\!=\!\frac{t_{cw}^2}{t_w}\sum_{n}
\chi_n (y) \chi_n(y') \zeta\left(
\frac{\epsilon-\epsilon_w}{t_w}-\lambda_n
\right),
\label{eq:sigmaepsilon}
\end{eqnarray}
where $X=X_{\mathrm{L}}$ or $X_{\mathrm{R}}$.
The size of matrices in (\ref{eq:grWB1}) is finite,
$(L_s+2L_c)L_y\times (L_s+2L_c)L_y$.
We use the tilde to denote that they are the Green's function
and the self-energy not for the original system but for the
wide band system.

In the semi-infinite regions $\cal W$ we take the wide band limit.
In this limit the self-energy $\tilde{\Sigma}^{(w)r}$ becomes
independent of $\epsilon$ and the non-zero elements of
$\tilde{\Sigma}^{(w)r}$ are equal to
\begin{eqnarray}
\hspace{-2mm}
 \tilde{\Sigma}^{(w)r}_{Xy,Xy'}
=t_{\ell}\sum_{n} \chi_n(y)\chi_n(y')
\zeta\left(
\frac{\mu}{t_{\ell}}-\lambda_n
\right).
\label{eq:sigmaleadWB}
\end{eqnarray}
This limit is obtained by taking the limit,
\begin{eqnarray}
t_w,t_{cw},\epsilon_w\rightarrow\infty,
\end{eqnarray}
while keeping
\begin{eqnarray}
t_{cw}^2/t_{w}&=&t_{\ell} \label{eq:ratio1},\\
\epsilon_w/t_w&=&-\mu/t_{\ell} \label{eq:ratio2}.
\end{eqnarray}
(When $\mu=0$, we do not need to introduce $\epsilon_w$.)

A similar idea to use an energy independent
self-energy can be seen in many papers, for example,
Refs.~\citen{tanaka:02} and \citen{jauho:94}.
Two important differences from previous works are:
\begin{itemize}
 \item {We keep non-interacting regions up to a length $L_c$
 on both sides of the sample to reduce artifacts of
 taking the wide band limit.}
 \item {The self-energy for the wide band region is chosen
 so that electrons at $\epsilon=\mu$
 are not scattered at the boundaries between ${\cal C}$ and ${\cal W}$.
 This makes it easier to reduce the artifacts.}
\end{itemize}

The self-energy (\ref{eq:sigmaleadWB}) in the wide band limit is
equal to the self-energy at $\epsilon=\mu$ for the
original system, i.e., the self-energy (\ref{eq:sigmaepsilon})
with $t_w=t_{cw}=t_{\ell}$ and $\epsilon_w=0$.
This means that electrons at the Fermi energy $\epsilon=\mu$ are not
scattered at the boundaries between the regions ${\cal C}$ and
${\cal W}$.
When electron-electron interaction is neglected, the Landauer
conductance for the wide band system is exactly the same as that for the
original Landauer system since the self-energy only at $\epsilon=\mu$ is
relevant for the conductance in non-interacting systems.
When we take account of the electron-electron interaction,
the conductances for the wide band system and for the original system
are no longer the same, because electrons
below the Fermi energy affect the motion of electrons at the Fermi
energy through the HF self-energy.
To reduce such artifacts of taking
the wide band limit on the calculated conductance, 
we keep non-interacting regions up to a length $L_c$
on both sides of the sample.
We expect that the artifacts of taking the wide band limit decrease
as $L_c$ increases, and they are finally removed in the limit
$L_c\rightarrow\infty$.

The absence of boundary scattering for electrons at $\epsilon=\mu$ is
important for the efficiency in removing the artifacts. If electrons
near $\epsilon=\mu$ are scattered strongly at the two boundaries,
resonant states due to the boundary scattering can be formed
near $\epsilon=\mu$. In this case, the Green's
function $\tilde{G}^r(\mu)$ at the Fermi energy shows larger fluctuation as
a function of $L_c$, and we need to simulate systems with
longer $L_c$ to remove the artifacts. On the other hand, our choice
minimize the boundary scattering of electrons near the Fermi energy,
that makes it easier to reduce the artifacts of taking the wide band
limit.

By taking the wide band limit, the self-energy
$\tilde{\Sigma}^{(w)r}$ becomes independent of $\epsilon$.
We define an effective Hamiltonian by
\begin{eqnarray}
 \tilde{H}^{{\rm (eff)}r}&=&H_t+
 \tilde{\Sigma}^{(w)r} +\tilde{\Sigma}^{({\rm HF})}.
\end{eqnarray}
Then the Green's function (\ref{eq:grWB1}) is written as
\begin{eqnarray}
  \tilde{G}^r(\epsilon)&=&
  \left[\epsilon-\tilde{H}^{({\rm eff})r}+\mathrm{i}\eta
  \right]^{-1}.
 \label{eq:grWB2}
\end{eqnarray}
Note that $\tilde{H}^{({\rm eff})r}$ is not a Hermitian matrix but is
a complex symmetric matrix since $H_t$ and $\tilde{\Sigma}^{\rm (HF)}$
are real symmetric matrices and $\tilde{\Sigma}^{(w)r}$ is a complex
symmetric matrix. 
The effective Hamiltonian has right and left eigenvectors
\begin{eqnarray}
 \tilde{H}^{({\rm eff})r} |r_n \rangle &=& q_n |r_n \rangle , \\
 \langle l_n | \tilde{H}^{({\rm eff})r} &=& \langle l_n |q_n .
\end{eqnarray}
Here $q_n$ is an eigenvalue, which is complex in general. Since the
effective Hamiltonian is a complex symmetric matrix, the transpose of
the corresponding right eigenvector is the left eigenvector,
\begin{eqnarray}
 \langle l_n | =|r_n\rangle ^{\top}.
\end{eqnarray}
For convenience, we impose the following normalization conditions.
\begin{equation}
\langle l_n | r_m \rangle = \delta_{n,m}.
\end{equation}
Then they satisfies the completeness relation,
\begin{equation}
\sum_n |r_n \rangle \langle l_n|=1.
\label{eq:completeness}
\end{equation}
By using eigenvalues $q_n$ and right eigenvectors $|r_n\rangle$,
the retarded Green's function is expressed as
\cite{datta,morse}
\begin{eqnarray}
\tilde{G}^r_{i,j}(\epsilon)
= \sum_n \frac{\phi_n(i)\phi_n(j)}
{\epsilon-a_n+\mathrm{i}b_n+\mathrm{i}\eta}.
\label{grij}
\end{eqnarray}
Here $a_n$ and $(-b_n)$ are the real part
and the imaginary part of the eigenvalue, $q_n=a_n-\mathrm{i}b_n$,
and $\phi_n(j)=\langle j|r_n \rangle$.

To calculate the self-energy $\tilde{\Sigma}^{\rm (HF)}$, we need to perform
the integral of the imaginary part of the retarded Green's function
\begin{equation}
 J_{i,j}(\mu,-\epsilon_c)=
 -\int_{-\epsilon_0}^{\mu} \mathrm{d}\epsilon
 {\rm Im} \tilde{G}^r_{i,j}(\epsilon).
 \label{eq:integral0}
\end{equation}
For a moment, we introduce a cutoff parameter $\epsilon_c$.
We will take the limit $\epsilon_c\rightarrow\infty$ later.
By using the expression (\ref{grij}),
the integral (\ref{eq:integral0}) can be performed analytically as
\begin{eqnarray}
 J_{i,j}(\mu,-\epsilon_c)
\!\!\!\!\!&=&\!\!\!\!\! \sum_n \Biggl\{
 {\rm Re}\left[\phi_n(i)\phi_n(j)\right]
 \left[\theta_n(\mu)-\theta_n(-\epsilon_c) \right]\nonumber \Biggr.\\
&& \!\!\!\!\!\Biggl.
 +{\rm Im}\left[\phi_n(i)\phi_n(j)\right]
 \ln \left[\frac{\cos \theta_n(\mu)}{\cos \theta_n(-\epsilon_c)}\right]
 \Biggr\}, \nonumber\\
\label{eq:integral}
\end{eqnarray}
where $\theta_n(\epsilon)$ ($\in[-\pi/2\,\pi/2]$) is defined by
\begin{eqnarray}
 \theta_n(\epsilon) = {\rm Tan}^{-1}\left(\frac{\epsilon-a_n}{b_n+\eta}\right).
\end{eqnarray}
For the second term in (\ref{eq:integral}),
we should not take the limit $\epsilon_c\rightarrow\infty$ before taking the
summation over $n$ because of the logarithmic divergence
\begin{eqnarray}
   \ln\left[\cos\theta_n(-\epsilon_c)\right]=-\ln \epsilon_c +\ln (b_n+\eta)
   +O(\epsilon_c^{-1}).
\end{eqnarray}
The logarithmic divergence disappears when we take the summation over $n$
because (\ref{eq:completeness}) implies
\begin{equation}
 \sum_n {\rm Im} \left[\phi_n(i)\phi_n(j)\right]=0.
\end{equation}
Therefore, in the limit $\epsilon_c\rightarrow\infty$ we have
\begin{eqnarray}
 J_{i,j}(\mu,-\infty)
 = \sum_n \Biggl\{
 {\rm Re}\left[\phi_n(i)\phi_n(j)\right]
 \left[\theta_n(\mu)+\frac{\pi}{2} \right]\nonumber \Biggr.\\
 \Biggl.
 +{\rm Im}\left[\phi_n(i)\phi_n(j)\right]
 \ln \left[\frac{\cos \theta_n(\mu)}{b_n+\eta}\right]
 \Biggr\}\label{eq:intfinal}.
\end{eqnarray}
Thus we calculate the self-energy from the eigenvalues and
the right eigenvectors of $\tilde{H}^{\rm (eff)}$ without
performing numerical integration.

Finally the numerical implementation of the wide band method is
summarized. First we prepare an initial matrix for
$\tilde{\Sigma}^{({\rm HF})}$. Then we calculate the eigenvalues
and eigenvectors by diagonalizing the effective Hamiltonian
$\tilde{H}^{({\rm eff})r}$. From the eigenvalues and right eigenvectors,
we calculate the self-energy $\tilde{\Sigma}^{({\rm HF})}$,
(\ref{eq:hartreeWB}) and (\ref{eq:exchangeWB}), by using
(\ref{eq:intfinal}). We continue this self-consistent iteration
until $\tilde{\Sigma}^{(\rm HF)}$ converges with enough precision.
(To ensure convergence, we have used ``the method of potential mixing''.
\cite{lee:02}) After convergence, we calculate the conductance
$\tilde{g}(L_c)$ for the wide band system by using the expression
(\ref{eq:landauerconductance}), where $\tilde{G}^{r,s}(\mu)$
is substituted for $G^{r,s}(\mu)$.
We expect that the conductance $\tilde{g}(L_c)$
for the wide band system approaches to the conductance
$g$ for the original system in the limit $L_c\rightarrow\infty$,
\begin{equation}
 g=\lim_{L_c\rightarrow\infty}\tilde{g}(L_c).
\end{equation}
In the case of 1D system (\S \ref{sec:result1d})
we extrapolate $\tilde{g}(L_c)$ by using an empirical fitting function.
In the case of 2D system (\S \ref{sec:result2d})
we make $L_c$ large enough so that the artifacts of the wide band limit
can be negligible to a good approximation.

\section{Application to a 1D system}
\label{sec:result1d}

\subsection{System}
Here we calculate the Landauer conductance in a 1D
system, i.e. $L_y=1$, described by the Hamiltonian (\ref{eq:horg}).
We suppose that the hopping parameter is uniform
in the system and set it unity as a unit of
energy, $t_s=t_{\ell}=t_{\ell s}=1$.
We also suppose that two electrons are interacting
only when they occupy the nearest neighbor sites,
\begin{equation}
 U_{x,x'}=
 \left\{
 \begin{array}{cc}
 U & {\rm (if}\hspace{1mm} x'=x\pm 1) \\
 0 & {\rm (otherwise)}.
 \end{array}
 \right.
\end{equation}
Transport phenomena in this model was previously studied in
Refs.~\citen{molina:03,molina:04-1,molina:04-2,meden:03}.
We set the positive background charge $K=0.5$ and the Fermi
energy $\mu=0$, which corresponds to half-filling.
Since we use the first order perturbation or the self-consistent
HF approximation the electron-electron interaction
should be weak, so we set $U=0.5$.

\subsection{In the first order perturbation}

\begin{figure}[tb]
\includegraphics[width=\linewidth]{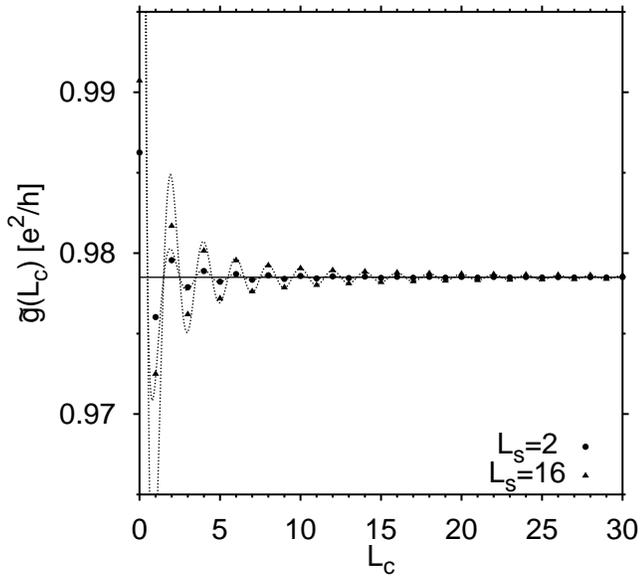}
\caption{The conductance $\tilde{g}(L_c)$ for the wide band system
calculated with the
wide band method in 1D in the first order perturbation is shown
as a function of $L_c$.
We set $K=0.5$, $\mu=0.0$, and $U=0.5$.
The dotted lines are the fit of
(\ref{eq:oscillation}) to the data with $L_c=[11,30]$.
The fit deviates from the numerical data when $L_c$ is much
smaller than this range.
The solid line indicates the conductance for the original system,
$g\approx 0.9785e^2/h$, obtained from (\ref{eq:1stexact}).}
\label{fig08:first}
\end{figure}

First we calculate the Landauer conductance in the first order
perturbation to test the wide band method.  Since the system
studied here is very simple, the conductance in the first order
perturbation for the original Landauer system (\ref{eq:horg})
can be calculated analytically without making use of the wide band
method. By comparing the numerical results of the wide band method
with the analytical results, we test the wide band method.

The self-energy in the first order perturbation for the original
system (\ref{eq:horg}) is obtained by replacing $G^r$ in
(\ref{eq:hartree}) and (\ref{eq:exchange}) with the Green's function
$G_0^r$ for the non-interacting Hamiltonian $H_0$. The first
order Hartree term is zero because the uniform negative charge of
the electrons cancels with the uniform positive charge in the background.
The first order exchange term at $\mu=0$ is given by
\begin{equation}
 \Sigma_{x,x+1}^{\rm (1st)}=-\frac{U}{\pi}
 \hspace{5mm}(1\leq x \leq L_s-1).
\end{equation}
From this self-energy, we obtain the conductance in the first order
perturbation. When $L_s$ is odd, we have a perfect transmission,
\begin{eqnarray}
 g=\frac{e^2}{h} \hspace{10mm}(L_s:{\rm odd}).
\end{eqnarray}
For even $L_s$ we have
\begin{eqnarray}
 g= \frac{e^2}{h}\left[\frac{2(1+U/\pi)}{(1+U/\pi)^2+1}
 \right]^2 \hspace{5mm}(L_s:{\rm even}).
 \label{eq:1stexact}
\end{eqnarray}
When $U=0.5$ and $L_s$ is even, we find $g \simeq 0.9785e^2/h$.

We have also calculated the conductance in the first order
perturbation by using the wide band method. The numerical result in
the first order perturbation has been obtained by stopping the
self-consistent iteration after just one iteration. When $L_s$ is
odd, the conductance is always $e^2/h$ independent of $L_c$ and
$L_s$. When $L_s$ is even, the conductance is reduced from $e^2/h$.
As shown in Fig.~\ref{fig08:first}, $\tilde{g}(L_c)$
oscillates as a function of $L_c$ when $L_s$ is even.

To remove the artifacts of taking the wide band limit, we need to
extrapolate the conductance $\tilde{g}(L_c)$ to $L_c\rightarrow\infty$.
The extrapolation has been done by using an empirical fitting function
of the form,
\begin{equation}
 \tilde{g}(L_c)=g+a\frac{\cos(\pi L_c)}{L_c^y}.
 \label{eq:oscillation}
\end{equation}
Here $g$, $a$, and $y$ are fitting parameters. We expect that the
asymptotic value $g$ is equal to the conductance for the original
system. By fitting numerical data with $L_c=[11,30]$,
we have found $g\approx 0.9785e^2/h$ for any even $L_s$ in the range
$L_s=[2,16]$. The estimates of $a$ and $y$ weakly depend on the range
of $L_c$ used for the fit. On the other hand, the estimate of
the asymptotic value $g$ is stable against the change of the range
of $L_c$. 

The asymptotic value $g\approx 0.9785e^2/h$ for even $L_s$ estimated
with the wide band method is in good agreement with the analytical
result (\ref{eq:1stexact}) for the original Landauer geometry.
The difference of the conductance is of order $10^{-8}$ to $10^{-6}$
in units of $e^2/h$.
The good agreement indicates that the wide band method works well
to estimate the conductance $g$ by extrapolating the conductance
$\tilde{g}(L_c)$ for the wide band system to $L_c\rightarrow\infty$.

\subsection{In the self-consistent Hartree-Fock approximation}

\begin{figure}[tb]
\includegraphics[width=\linewidth]{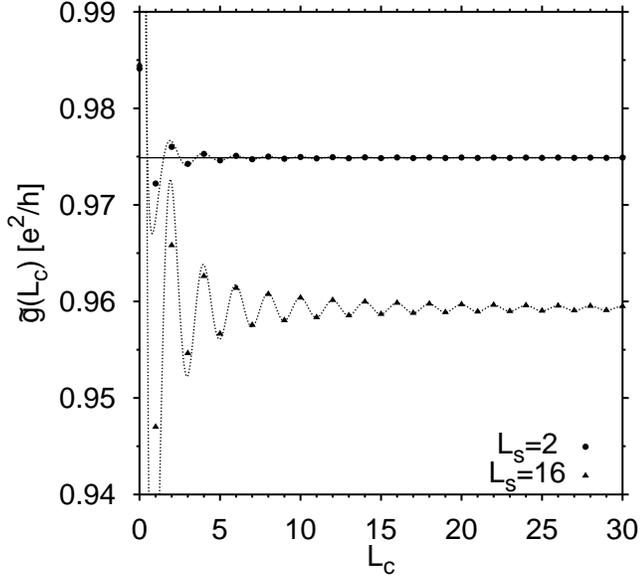}
\caption{The conductance $\tilde{g}(L_c)$ for the wide band system
calculated with the
wide band method in 1D in the HF approximation is shown
as a function of $L_c$.
We set $K=0.5$, $\mu=0.0$, and $U=0.5$.
The dotted lines are the fit
of (\ref{eq:oscillation}) to the data with $L_c=[11,30]$.
The fit deviates from the numerical data when $L_c$ is much smaller
than this range. The solid line indicates the conductance for the
original system with $L_s=2$, $g\approx 0.9749e^2/h$,
obtained from (\ref{eq:1dhfequation}) and (\ref{eq:1dconductanceLs2}).
There is no solid line for $L_s=16$ since it is too difficult for us
to calculate the conductance for the original system with $L_s=16$
in the HF approximation.
}
\label{fig:hflc}
\end{figure}

We have then calculated the conductance in 1D
in the HF approximation with the wide band method.
When $L_s$ is odd, we have found $\tilde{g}(L_c)=e^2/h$ for any $L_c$.
When $L_s$ is even, the conductance is reduced from $e^2/h$.
We extrapolated the conductance for each even $L_s$ to
$L_c\rightarrow\infty$ by using the fitting function (\ref{eq:oscillation}),
as in the case of the first order perturbation. Some numerical data
and the corresponding fits are shown in Fig.~\ref{fig:hflc}. From the
fit, we have obtained the asymptotic value $g$ for each even $L_s$.

Even in the HF approximation it is not impossible to calculate the
conductance $g$ for the original system without using the
wide band method. So far we have calculated $g$ only for $L_s=2$.
The Hartree term $\Sigma_{x,x}^{\rm (HF)}$ is
zero when $\mu=0$ because of the particle-hole symmetry.
The exchange self-energy $\Sigma_{1,2}^{\rm (HF)}$ for $L_s=2$ is
a solution of
\cite{asada-pichard:unpub}
\begin{eqnarray}
 1 - v  = -U \!
 \left\{ \frac{v^2 \! - \! 1}{2v^2}
\! \left[1 - \frac{1}{\pi}{\rm Tan}^{-1} \!\!
\left(\!\frac{2v}{v^2\! -\! 1}\!\right)
\! \right]  +  \frac{1}{\pi v} \right\},
\label{eq:1dhfequation}
\end{eqnarray}
with $v=1-\Sigma_{1,2}^{\rm (HF)}$. From this equation
we find $\Sigma_{1,2}^{\rm (HF)}\approx -0.1733$ for $U=0.5$.
By using a formula for the conductance
\begin{eqnarray}
 g=\frac{e^2}{h}\left[\frac{2(1-\Sigma_{1,2}^{\rm (HF)})}
 {(1-\Sigma_{1,2}^{\rm (HF)})^2+1}\right]^2,
 \label{eq:1dconductanceLs2}
\end{eqnarray}
we find $g\approx 0.9749e^2/h$ for $L_s=2$ and $U=0.5$.
This value of the conductance is indicated with a solid line
in Fig.~\ref{fig:hflc}.
The difference between this value and that
estimated with the wide band method is of order $10^{-8}e^2/h$,
indicating that the wide band method works well.

\subsection{Length dependence of the conductance --
comparison of numerical results}

\begin{figure}[tb]
\includegraphics[width=\linewidth]{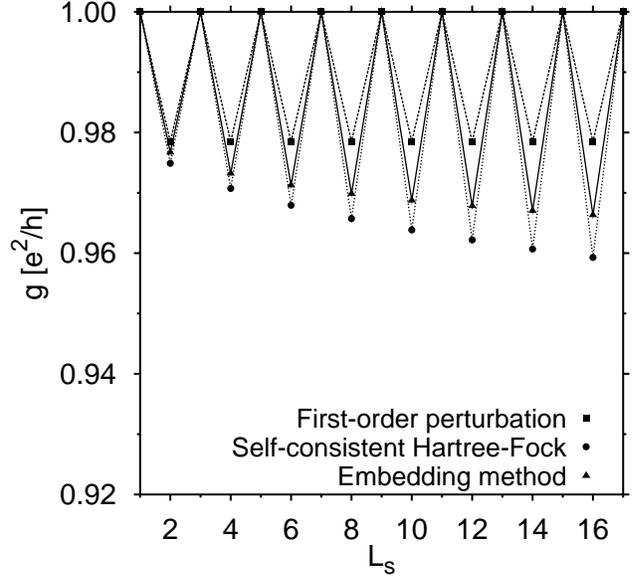}
\caption{$L_s$ dependent oscillation of the conductance $g$ in 1D.
The conductance $g$ in the first order perturbation and
in the HF approximation are shown as well as that of the
embedding method. \cite{molina:04-2}
We set $K=0.5$, $\mu=0.0$, and $U=0.5$.
The data calculated by the embedding method
were provided by R. A. Molina and J.-L. Pichard.
\cite{molina:unpub}
The lines are a guide to the eye only.
}
\label{fig:1dls}
\end{figure}

Figure \ref{fig:1dls} shows the $L_s$ dependence of the conductance
within the first order perturbation and within the HF approximation.
For a reference, the conductance calculated with the embedding method
by R. A. Molina and J.-L. Pichard \cite{molina:unpub} is also shown.
In the embedding method, the electron-electron interaction is
treated exactly, hence it is thought that the calculated conductance
is also exact.

In all three cases, the perfect conductance $g=e^2/h$ is obtained
when $L_s$ is odd, and the conductance is reduced from it
when $L_s$ is even. This even-odd oscillation was found
in Ref.~\citen{molina:04-2}. A similar parity oscillation
in the Hubbard chain was reported in Ref.~\citen{oguri:99}.

Within the first order perturbation,
the conductance is independent of $L_s$ for all even $L_s$. However,
the result of the embedding method, which is thought to be exact,
indicates that the conductance for even $L_s$ decreases when $L_s$
increases. The conductance for even $L_s$ in the HF approximation
also shows that the conductance decreases with increasing $L_s$. So
we find a qualitative agreement
between the behavior of the conductance
in the HF approximation and that of the embedding method.
Furthermore, the results indicate that the HF approximation is more
accurate than the first order perturbation quantitatively.

\section{Application to a 2D system}
\label{sec:result2d}

\subsection{System}
The application of the
wide band method is not restricted to 1D systems.
Here we apply it to a 2D system, i.e., a square system ($L_s=L_y$).
The hopping parameter is supposed to be uniform and set it unity
as a unit of energy, $t_s=t_{\ell}=t_{\ell s}=1$.
We consider the nearest neighbor electron-electron interaction
in the sample region,
\begin{equation}
 U_{i,j}=
 \left\{
 \begin{array}{cc}
 U & {\rm (if}\hspace{1mm} (i,j)\hspace{1mm}
 {\rm is}\hspace{1mm} {\rm a} \hspace{1mm} {\rm n.n}
 \hspace{1mm}{\rm pair)} \\
 0 & {\rm (otherwise)}.
 \end{array}
 \right.
\end{equation}
In our simulation, we set the strength of electron-electron
interaction $U=0.5$. We set the chemical potential $\mu=0$
and the positive charge $K=0.5$, which corresponds to half-filling.

\subsection{In the first order perturbation}
\begin{figure}[tb]
\includegraphics[width=\linewidth]{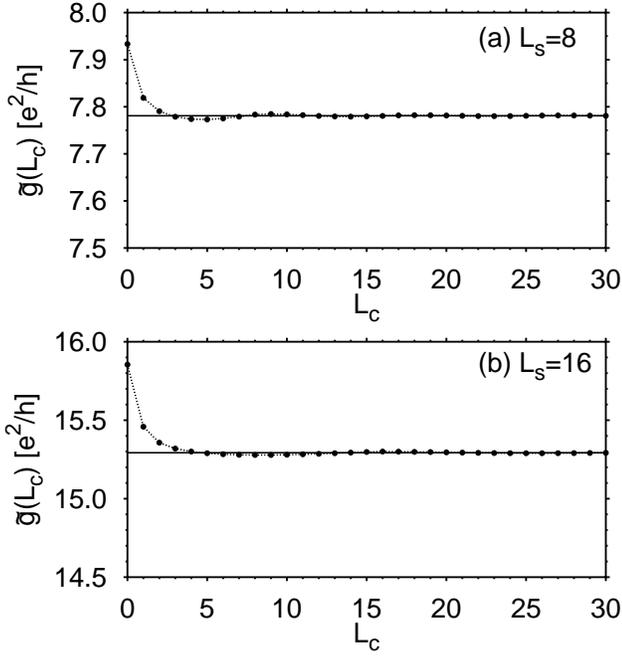}
\caption{The conductance $\tilde{g}(L_c)$ for the wide band system
calculated with the
wide band method in 2D in the first order perturbation 
for (a)$L_s=8$ and (b)$L_s=16$ is shown as a function of $L_c$.
We set $K=0.5$, $\mu=0.0$, and $U=0.5$.
The dotted lines are a guide to the eye only.
The solid lines indicates the conductance $g$ for the original
system obtained from (\ref{eq:exchange2d-1}) and (\ref{eq:exchange2d-2})
without using the wide band method.
}
\label{fig:fo2dlc}
\end{figure}
First we have calculated the conductance within the first order
perturbation to test the wide band method in 2D as it has been
tested in 1D in \S \ref{sec:result1d}.

Within the first order perturbation, we can calculate the
conductance without using the wide band method since the
self-energy $\Sigma^{\rm (1st)}$ in the first order
perturbation for the original Landauer system is obtained
as follows. The Hartree contribution is zero since the positive
background charge compensate the negative charge of electrons.
The exchange contribution is given by
\begin{eqnarray}
 \Sigma^{({\rm 1st})}_{x,y;x+1,y}=- \frac{2U}{(L_y+1)\pi}
 \sum_{n=1}^{L_s} \sin k_y^{(n)}
 \sin^2\left(k_y^{(n)}y\right),
 \label{eq:exchange2d-1}
\end{eqnarray}
\begin{eqnarray}
 \Sigma^{({\rm 1st})}_{x,y;x,y+1}&=& \frac{2U}{(L_y+1)\pi}
 \sum_{n=1}^{L_s} k_y^{(n)}\sin \left(k_y^{(n)}y\right)
 \nonumber \\
 &&\times 
 \sin\left[k_y^{(n)}(y+1)\right],
 \label{eq:exchange2d-2}
\end{eqnarray}
where $k_y^{(n)}=\pi n/(L_s+1)$.
We have calculated $\Sigma^{\rm (1st)}$ by taking the summation over
$n$ in (\ref{eq:exchange2d-1}) and (\ref{eq:exchange2d-2}) numerically,
and then have calculated the corresponding conductance $g$.

We have also calculated the conductance making use of the wide band
method in the first order perturbation. Figure~\ref{fig:fo2dlc} shows
the $L_c$ dependence of
the conductance $\tilde{g}(L_c)$ for the wide band systems with
sample size $L_s=8$ and $L_s=16$. For a reference, the conductance $g$
calculated from (\ref{eq:exchange2d-1}) and (\ref{eq:exchange2d-2})
is also shown with the solid line. The figure indicates that
$\tilde{g}(L_c)$ fluctuates around the value $g$
and the fluctuation becomes smaller when $L_c$ increases.
It is reasonable to assume that
the influence of the wide band limit is, to a good approximation,
negligible when $L_c=30$,
\begin{equation}
 g\approx \tilde{g}(L_c=30).
\end{equation}
We have found that the differences between the value $g$ obtained from
(\ref{eq:exchange2d-1}) and (\ref{eq:exchange2d-2})
and $\tilde{g}(L_c=30)$ are of order $10^{-5}$ to $10^{-3}$ in units of
$e^2/h$ for $L_s=[2,16]$. This indicates that the wide band method also
works well in 2D. 

\subsection{In the self-consistent Hartree-Fock approximation}

\begin{figure}[tb]
\includegraphics[width=\linewidth]{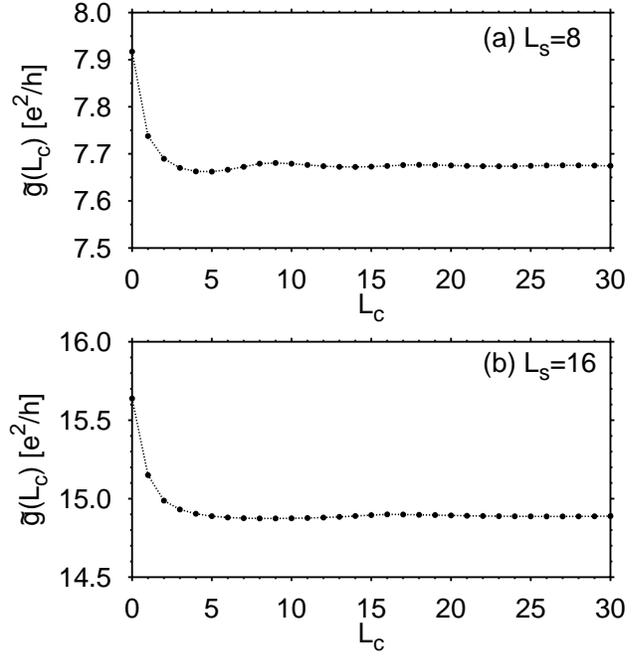}
\caption{The conductance $\tilde{g}(L_c)$ for the wide band system
calculated with the
wide band method in 2D in the HF approximation for (a)$L_s=8$
and (b)$L_s=16$ is shown as a function of $L_c$.
We set $K=0.5$, $\mu=0.0$, and $U=0.5$.
The dotted lines are a guide to the eye only.
}
\label{fig:hf2dlc}
\end{figure}
We have then calculated the conductance in 2D with the wide band method
in the HF approximation. Figure~\ref{fig:hf2dlc} shows the $L_c$
dependence of the conductance $\tilde{g}(L_c)$ for the wide band systems
with sample size $L_s=8$ and $L_s=16$. The $L_c$ dependence of
$\tilde{g}(L_c)$ is qualitatively similar to that in the first order
perturbation. We again assume that the influence of the wide band limit
is negligible when $L_c=30$.

In the range $L_c=[21,30]$, the fluctuation of $\tilde{g}(L_c)$ is
of order $10^{-3}$ in units of $e^2/h$. Assuming that $\tilde{g}(L_c)$ is
fluctuating around the asymptotic value $g$, we can consider the
amplitude of the fluctuation as a precision of $g$ in the wide band
method.

\subsection{$L_s$ dependence of the conductance}

\begin{figure}[tb]
\includegraphics[width=\linewidth]{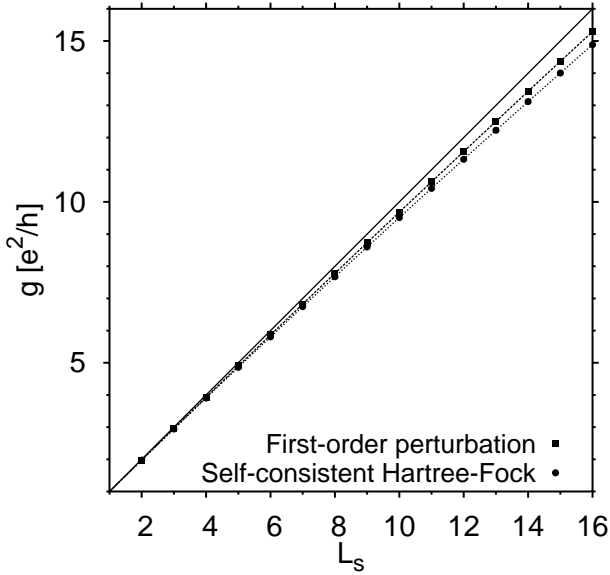}
\caption{$L_s$ dependence of the conductance $g$ in 2D
in the first-order perturbation and in the
HF approximations.
We set $K=0.5$, $\mu=0.0$, and $U=0.5$.
For a reference, the conductance
corresponding to the perfect transmission $N_c e^2/h$ is also shown.
(At $\mu=0$, $N_c=L_s$ in the 2D strip imposed fixed boundary conditions
in the transverse direction.)
The dotted and dashed lines are a guide to the eye only.
}
\label{fig:2dls}
\end{figure}

Figure \ref{fig:2dls} shows $L_s$ dependence of the conductance
in the first order perturbation and in the HF approximation in 2D.
For a reference, the value $N_c e^2/h$ corresponding to
a perfect transmission,
with $N_c$ being the number of propagating channels,
is also shown.
The conductance in the HF approximation
tends to be smaller than that in the first order perturbation as it is
in 1D

\section{Summary and Discussion}
\label{sec:summary}
We have developed a new numerical method
to calculate the Landauer conductance through an interacting electron
system at zero temperature in the first order perturbation
or in the self-consistent HF approximation.
A troublesome numerical integration is avoided by taking a wide band
limit. We can remove the artifacts of taking the wide band limit by
increasing the length of non-interacting region $L_c$ kept on both sides
of the sample. We have applied it to 1D and 2D interacting systems.

The method does not require much CPU time, so it permits us to
accumulate many samples in studying interacting disordered systems.
Simulation in the presence of disorder is left for future.

The wide band method has an advantage that the dimensionality
is not restricted to one. This method can be useful when
studying various quantum transport phenomena not only in
interacting disordered systems
but also in quantum dots, quantum point contacts, quantum nanowires,
atomic chains, and so on.
It is possible
to consider spin degree of freedom.
It is also possible to take account of other effects,
such as spin-orbit coupling and a magnetic field,
by generalizing the wide band method. In the presence of such effects,
the effective Hamiltonian might be no longer a complex symmetric matrix.
When it is not a complex symmetric matrix,
we need to calculate not only the right eigenvectors
but also the left eigenvectors.

\begin{acknowledgments}
The author thanks Professor Keith Slevin for suggesting him to perform
a numerical simulation in the HF approximation,
for his critical reading of the manuscript,
and for valuable discussions.
He would like to thank Professor Jean-Louis Pichard and
Dr. Rafael A. Molina for suggesting to put non-interacting regions
on the sample, for providing the numerical data of the embedding
method used in Fig.~\ref{fig:1dls}, and for stimulating discussions.
He thanks Professor Tomi Ohtsuki and Mr. Axel B. Freyn for valuable
discussions. He acknowledges the support of the 21st
century COE program of Osaka University
``Towards a new basic science; depth and synthesis''.
He is grateful to Research Fellowships of the Japan Society
for the Promotion of Science for Young Scientists.
\end{acknowledgments}

\appendix
\section{Green's functions for semi-infinite leads and some
related matrices}
\label{app:green}
In this appendix, the retarded Green's function
for semi-infinite leads and some related matrices are shown
explicitly.
We suppose that the leads are described by the Hamiltonian
(\ref{eq:hl}) and coupled with a sample by (\ref{eq:hls}).

First we calculate the retarded Green's function $g^r(\epsilon)$
for the isolated lead at the left ($x\leq 0$). In particular,
the important elements are those at the surface of the lead, i.e.,
the elements $g_{0y,0y'}^r(\epsilon)$.
The eigenfunctions and eigenvalues of the lead are written as
\begin{eqnarray}
 \psi_{k_x,n}(x,y)&=&\sqrt{\frac{2}{\pi}} \sin \left[k_x (x-1)\right]
 \chi_n(y)\label{eq:psikxn}, \\
 E_{k_x,n}&=&-2t_{\ell} \cos k_x +t_{\ell}\lambda_n.
 \label{eq:ekxn}
\end{eqnarray}
where
\begin{eqnarray}
\chi_n(y)&=&\frac{1}{\sqrt{2(L_y+1)}}\sin \left(k_y^{(n)} y\right), \\
\lambda_n&=&-2 \cos k_y^{(n)}.
\end{eqnarray}
Here $k_x$ ($\in[0,\pi]$) is the wave number in the $x$-direction,
and $k_y^{(n)}=\pi n/(L_y+1)$.
Using them we have
\begin{eqnarray}
g_{0y,0y'}^r(\epsilon)= 
\frac{1}{t_{\ell}}\sum_{n=1}^{L_y}
\chi_n(y)\chi_n(y')
\zeta\left(\frac{\epsilon}{t_{\ell}}-\lambda_n\right).
\label{eq:glr}
\end{eqnarray}
Here the function $\zeta(z_n)$ is given by
\begin{eqnarray}
\zeta(z_n)
=\left\{
\begin{array}{cc}
\frac{z_n}{2}+\sqrt{\frac{z_n^2}{4}-1}&(z_n<-2)\\
\frac{z_n}{2}-\mathrm{i} \sqrt{1-\frac{z_n^2}{4}}&(-2\leq z_n\leq 2)\\
\frac{z_n}{2}-\sqrt{\frac{z_n^2}{4}-1}&(z_n>2).
\end{array}
\label{eq:fr}
\right.
\end{eqnarray}
Similarly,
the Green's function for the other lead at the right ($x\geq L_s+1$)
is obtained as
\begin{eqnarray}
g_{L_s+1y,L_s+1y'}^r(\epsilon)= 
\frac{1}{t_{\ell}}\sum_{n=1}^{L_y}
\chi_n(y)\chi_n(y')
\zeta\left(\frac{\epsilon}{t_{\ell}}-\lambda_n\right).
\label{eq:grr}
\end{eqnarray}

A transverse mode $n$ is a propagating mode if 
$|\mu/t_{\ell}-\lambda_n|\leq 2$
and an evanescent mode otherwise.
The Fermi wave number $k_F^{(n)}$ ($0\leq k_n \leq \pi$)
of the propagating mode $n$ in the $x$-direction
is determined by
\begin{equation}
 \mathrm{e}^{-\mathrm{i} k_F^{(n)}}=\frac{z_n}{2}-\mathrm{i}
\sqrt{1-\frac{z_n^2}{4}},
\end{equation}
with $z_n=\mu/t_{\ell}-\lambda_n$.

When calculating the Landauer conductance
the semi-infinite leads and the sample are coupled
by (\ref{eq:hls}).
In the Green's function method
the effects of the lead are taken account
in terms of the self-energy. \cite{datta,ando:91}
The retarded self-energy $\Sigma_{xy,x'y'}^{(\ell)r}$ due to the leads
becomes non-zero at the surfaces of the sample, i.e.,
when $x=x'=1$ or $x=x'=L_s$.
The non-zero elements are defined by
$\Sigma^{(\ell)r}_{1y,1y'}(\epsilon)=t_{\ell s}^2 g_{0y,0y'}^r(\epsilon)$
and $\Sigma^{(\ell)r}_{L_sy,L_sy'}(\epsilon)=t_{\ell s}^2
g_{L_s+1y,L_s+1y'}^r(\epsilon)$.
Using the expression (\ref{eq:glr}) we have
\begin{eqnarray}
 \Sigma^{(\ell)r}_{Xy,Xy'}(\epsilon)=
\frac{t_{\ell s}^2}{t_{\ell}}\sum_{n=1}^{L_y}
\chi_n(y)\chi_n(y')
\zeta\left(\frac{\epsilon}{t_{\ell}}-\lambda_n\right),
 \label{eq:sigmaLr}
\end{eqnarray}
where $X=1$ or $L_s$.

The $L_y\times L_y$ matrices $\Gamma^{\rm (L)}$ and
$\Gamma^{\rm (R)}$, which appear in the Landauer formula
(\ref{eq:landauerconductance}), are defined by
$\Gamma_{yy'}^{\rm (L)}(\epsilon)= \mathrm{i}
\left[\Sigma_{1y,1y'}^r(\epsilon)
-\Sigma_{1y,1y'}^a(\epsilon)\right]$ and
$\Gamma_{yy'}^{\rm (R)}(\epsilon)= \mathrm{i}
\left[\Sigma_{L_s y,L_s y'}^r(\epsilon)
-\Sigma_{L_s y,L_s y'}^a(\epsilon)\right]$.
At $\epsilon=\mu$ we have
\begin{eqnarray}
  \Gamma_{yy'}^{\rm (L,R)}(\mu)
  =\frac{2t_{\ell s}^2}{t_{\ell}}{\sum_{n}}^{\prime}
  \chi_n(y) \chi_n(y') \sin k_F^{(n)}.
\end{eqnarray}
Here the summation is taken over propagating modes.

%\bibliography{wideband}

\end{document}